\documentclass[11pt]{article} % use larger type; default would be 10pt

\usepackage[utf8]{inputenc} % set input encoding (not needed with XeLaTeX)
\usepackage[english]{babel}

\usepackage{geometry} % to change the page dimensions
\usepackage{graphicx} % support the \includegraphics command and options

\usepackage{booktabs} % for much better looking tables
\usepackage{array} % for better arrays (eg matrices) in maths
\usepackage{paralist} % very flexible & customisable lists (eg. enumerate/itemize, etc.)
\usepackage{verbatim} % adds environment for commenting out blocks of text & for better verbatim
\usepackage{subfig} % make it possible to include more than one captioned figure/table in a single float
\usepackage{natbib} % bibliography
\usepackage{amsmath}
\usepackage{amssymb} % for fancy math symbols
\usepackage{fancyhdr} % This should be set AFTER setting up the page geometry
\usepackage{sectsty}
\allsectionsfont{\sffamily\mdseries\upshape} % (See the fntguide.pdf for font help)
\usepackage[nottoc,notlof,notlot]{tocbibind} % Put the bibliography in the ToC
\usepackage[titles,subfigure]{tocloft} % Alter the style of the Table of Contents

\title{Quantifying Population Movement Using a Novel Implementation of Digital Image Correlation in the ICvectorfields package}
\author{Devin W. Goodsman}
\begin{document}
\bibliographystyle{plainnat}
\maketitle

\section*{Abstract}

Movements in imagery captivate the human eye and imagination. They are also of interest in variety of scientific disciplines that study spatiotemporal dynamics. Popular methods for quantifying movement in imagery include particle image velocimetry and digital image correlation. Both methods are widely applied in engineering and materials science, but less applied in other disciplines. This paper describes an implementation of a basic digital image correlation algorithm in R \citep{R} as well as an extension designed to quantify persistent movement velocities in sequences of three or more images. Algorithms are applied in the novel arena of landscape ecology to quantify population movement and to produce vector fields for easy visualization of complex movement patterns across space. Functions to facilitate analyses are available in \textbf{ICvectorfields} \citep{ICvf}. These methods and functions are likely to produce novel insights in theoretical and landscape ecology because they facilitate visualization and comparison of theoretical and observed data in complex and heterogeneous environments.

\section*{Introduction}

Living organisms move through space in complex ways that have inspired many branches of spatial pattern analysis from Turing instabilities \citep{Alonso2002, Ruan1998}, to complex systems analysis of the emergent properties of individual-level behaviours when organisms live in groups \citep{Parrish1999, Johnson2009}. Moreover, in mathematical ecology there is a long history of deriving analytic expressions for traveling wave speeds from mathematical models of biological systems based on partial differential equations \citep{Kolmogorov1937, Skellam1951} and integrodifference equations \citep{Kot1996}. In addition to standard traveling waves and wave-trains, simulation studies have revealed more unusual patterns of population level movement can arise from the way organism interact and move on the landscape \citep{Hassell1994}. Spiral waves are one example of surprising spatiotemporal dynamics that can arise in biological systems \citep{Hassell1991}.

Travelling waves and spiral waves emerge from mathematical models of population expansion, which are often based on partial differential equations and integrodifference equations. These types of models, which represent the movement patterns of populations of organisms, are sometimes classified as Eulerian approaches to distinguish them from Lagrangian approaches that focus on the trajectories of individuals. The majority of R packages that quantify organismal movement, however, are Lagrangian as they pertain to the analysis of the tracks or trajectories of individual animals with tracking collars or tracking devices. Integrated step selection models \citep{Avgar}, such as those in the \textbf{amt} package \citep{amt}, which incorporate the impact of spatially variable habitat or environmental variables on movement of individuals modeled using a discrete time and discrete space framework, are an example of a Lagrangian approach when fitted to movement data from individuals. Because my focus in this work is on population-level movements that are evident in imagery, I will forego further discussion of Lagrangian models and instead refer the interested reader to a review of R packages for modeling animal movement \citep{Joo}. At the time of writing, R packages that focus on the Eulerian approach include \textbf{IDE} \citep{IDE}, \textbf{deSolve} \citep{deSolve}, and \textbf{ReacTran} \citep{ReacTran}. These packages are designed primarily to obtain numerical solutions to Eulerian models, analyze their dynamics, and fit them to data. Recently, movement modeling based on stochastic differential equations, stochastic partial differential equations \citep{Krainski}, and other stochastic process models \citep{Buderman} has proliferated. Computationally efficient Bayesian statistical approaches are often required to fit these stochastic models to data due to the ubiquity of noise in spatiotemporal time series in combination with nonlinear dynamical processes \citep{Krainski}. 

In contrast to the R packages and approaches I have cited above, this work is focused on the description of an empirical method for quantifying spatially heterogeneous rates of movement in sequences of images without fitting a model--although I do approach the quantification problem from an Eulerian perspective. Empirically quantifying spread rates without imposing a specific mathematical model allows the user to abandon many of the assumptions implicit in mathematical modeling of population expansion. For example, tractable mathematical models of consumer-resource systems that generate traveling waves, wave-trains, and spiral waves, often rely on assumptions of a homogeneous spatial environment with respect to resources or one in which there are no discontinuities in resources. In contrast, many organisms spread in environments that are spatially heterogeneous \citep{Urban2008}, and in environments subject to persistent directional flows that impact organism movement \citep{Hoffman2006}. The ramifications of this claim are more easily understood using a meteorological analogy. In meteorology, vector fields are frequently used to illustrate the impacts of high and low pressure systems on wind speed and direction, and thus on the movement of weather systems. In such meteorological systems, wind speed and direction are complicated functions of topography and complex atmospheric dynamics. As a result, vector fields representing movement in such systems are often variable at the regional scale, with winds flowing in one direction on one side of a map and possibly in an opposing direction on the other side. In ecology, populations of organisms are like the weather systems in that their movement on the landscape is what is of primary interest to researchers; variable wind causes spatially variable movement of weather systems similarly to how persistent directional fluid flows, including wind, in an organism's environment impact dispersal, and therefore population movement. 

At the time of writing, the only tool in R \citep{R} designed to empirically estimate spreading speed or the speed of wave-trains in populations without fitting a mathematical model is implemented in the \textbf{ncf} R package \citep{ncf}. The \textbf{ncf} package relies on lagged non-parametric cross-correlation functions to estimate spreading speed of traveling waves \citep{Bjornstad2001}. To do so, it takes two spatiotemporal data sets that differ from one another only in that one is a time-lagged version of the other, and projects their planar coordinates onto lines of varying angles that can be specified using function arguments. After projection onto a line, cross-correlation is estimated using a spline-correlogram approach \citep{Bjornstad2001} and the location of maximum cross-correlation gives an estimate of displacement along the direction of the projection line. This approach was used to estimate the velocity of traveling wave-trains in the larch budmoth system in the European Alps \citep{Bjornstad2002}. 

Projecting population data from a domain with two spatial coordinates onto a domain that has only one spatial coordinate and then using a correlogram approach precludes quantification of more complicated patterns of movement on the landscape. For example if two spatially separated populations are moving towards one another at the same speed, such an approach will yield enigmatic correlograms. Similarly, if several populations move radially around a central fulcrum, the correlogram will be difficult to translate to an estimate of directional movement.

In this paper I present an approach for estimating vector-fields in systems with spatially variable movement that is inspired by a technique from engineering and materials science called Digital Image Correlation or DIC \citep{Anuta1970, Sutton2009}. Among other things, Digital Image Correlation is used to estimate displacement based on photographs of a planar material before, during, and after a force has been applied to warp its surface \citep{Sutton2009}. A typical DIC approach as well as the extensions described in this paper are implemented in the \textbf{ICvectorfields} package \citep{ICvf}, in which the IC is the abbreviation for Image Correlation. I demonstrate these approaches using the \textbf{ICvectorfieds} package to analyze a simulated data set as well as the larch budmoth data set provided with the \textbf{ncf} R package \citep{ncf, Bjornstad2002}.

\section*{Mathematical and Computational Details}

Here I provide mathematical and computational details of the algorithms used in the \textbf{ICvectorfields} R package starting with a standard digital image correlation approach, and following with extensions to estimate persistent movement and to account for spatial variability in persistent movement. The \textbf{ICvectorfields} package capitalizes on the algorithms written in C under the title FFTW which stands for Fastest Fourier Transform in the West \citep{FFTW2005}, and a convenient wrapper package in R called \textbf{fftwtools} \citep{fftwtools}. Input raster images and raster stacks are read and manipulated using the \textbf{terra} package \citep{terra}.

\subsection*{Digital image correlation}

One of the earliest applications of cross-correlation in image analysis was to align images taken from different sensors or at different times using satellites or aircraft \citep{Anuta1970}. The theoretical and computational details I present here loosely follow those in this pioneering application. I will provide the mathematical underpinning of two-dimensional cross-correlation, and then elaborate on its computational implementation, which involves some additional complexity due to the circular nature of discrete fast Fourier transforms. In all descriptions below, I do not normalize the cross-correlation function to obtain Pearson correlation coefficients and therefore, I follow the convention of using the terms cross-correlation and cross-covariance interchangeably.

Given two images that have been converted to square matrices $f$ and $g$ of dimension $m \times m$, two-dimensional cross-correlation can be defined in terms of a convolution:

\begin{equation}
\left(f \star g\right)\left(x_j, y_i\right)  =  \left(\overline{f\left(-x_j, -y_i\right)} * g\left(x_j, y_i\right)\right)\left(x_j, y_i\right), \label{eq:1}
\end{equation}

\noindent in which $\left(f \star g\right)$ is the two-dimensional cross-correlation matrix, the $*$ operator denotes convolution, $\overline{f\left(-x_j, -y_i\right)}$ is the complex conjugate of the $f\left(x_j, y_i\right)$ matrix, $i$ is the matrix row index, and $j$ is the matrix column index $i, j \in \mathbb{N} = \{1, 2, . . .\}$. Note that I use array indices that start at one rather than zero. The coordinates of the centroids of each pixel are given by $x_j$ and $y_i$.

Based on the convolution theorem, equation \eqref{eq:1} can be rewritten as

\begin{equation}
\left(f \star g\right)\left(x_j, y_i\right)  = \mathbb{F}^{-1}\left(\overline{\mathbb{F}\left(f\left(x_j, y_i\right)\right)}\mathbb{F}\left(g\left(x_j, y_i\right)\right)\right)\left(x_j, y_i\right), \label{eq:2}
\end{equation} 

\noindent wherein $\mathbb{F}$ denotes the two-dimensional Fourier transform, $\mathbb{F}^{-1}$ denotes its inverse, and $\overline{\mathbb{F}\left(f\left(x_j, y_i\right)\right)}$ is the complex conjugate of $\mathbb{F}\left(f\left(x_j, y_i\right)\right)$. Because $\overline{\mathbb{F}\left(f\left(x_j, y_i\right)\right)} = \mathbb{F}\left(\overline{f\left(-x_j, -y_i\right)}\right)$, and because $f\left(x_j, y_i\right)$ contains only real numbers, I forego calculating the complex conjugate of the Fourier transform of the $f$ matrix and choose instead capitalize on matrix multiplication to compute the complex conjugate:

\begin{equation}
\overline{f\left(-x_j, -y_i\right)} = r  \times f \times r, \label{eq:3}
\end{equation}

\noindent in which the r matrix is a $m \times m$ matrix that has zeros everywhere except for along the diagonal that runs from its lower left to upper right corners, which contains ones: 

\begin{equation}
r = 
\begin{pmatrix}
0 & 0 & \cdots & 0 & 1 \\
0 & 0 & \cdots & 1 & 0 \\
\vdots  & \vdots  & \ddots & \vdots & \vdots \\
0 & 1 & \cdots & 0 & 0\\ 
1 & 0 & \cdots & 0 & 0
\end{pmatrix}. \label{eq:4}
\end{equation}

\noindent The matrix calculations in equations \eqref{eq:3} and \eqref{eq:4} substitute for calculation of the complex conjugate only in the case where all values of the $f$ matrix are real, which is the case in most natural science applications.

Together, equations \eqref{eq:1} through \eqref{eq:4} constitute an elegant way to compute two dimensional cross-correlation. Computer implementation of these, however, requires some additional complexity due to the use of discrete fast Fourier transforms to efficiently compute convolutions. Discrete fast Fourier transforms are inherently circular, which means that what happens on the outer edges of matrices will impact their discrete fast Fourier transform on the opposite side. In order to mitigate this problem, zeros are added to the outer edge on all sides of both the $f$ and $g$ matrices \citep{Anuta1970}. In \textbf{ICvectorfields}, the $f$ and $g$ matrices are padded with as many zeros as there are rows and columns in the original matrix and then additional zeros are added to ensure that both matrices are square with an even number of rows and columns. 

When matrices are padded as described above and discrete fast Fourier transforms are used as in equation \eqref{eq:2}, the non-cyclic components of the convolution end up in the outer corners of the cross-correlation matrix $\left(f \star g\right)\left(x_j, y_i\right)$ \citep{Anuta1970}. Thus, to obtain a correct estimate of cross-correlation, $\left(f \star g\right)\left(x_j, y_i\right)$ must be divided into four quadrants and each quadrant must be flipped along its horizontal and vertical axes using matrix multiplication. For example, if the zero-padded matrices have dimension $k \times k$, each quadrant of $\left(f \star g\right)\left(x_j, y_i\right)$ will have dimension $k/2 \times k/2$ and the following operation flips each quadrant matrix:

\begin{equation}
q_f = r \times q \times r, \label{eq:5}
\end{equation}

\noindent where $r$ is a $k/2 \times k/2$ matrix as in equation \eqref{eq:4}. Then the four quadrants can be reassembled into the $k \times k$ cross-correlation matrix. The mathematical operations in equations \eqref{eq:1} through \eqref{eq:5} are implemented in the \emph{Xcov2D} function in the \textbf{ICvectorfields} package.

Once a cross-correlation matrix has been estimated, it can be used to compute displacement in the horizontal and vertical directions in terms of the horizontal and vertical shifts in pixel numbers that maximize cross-correlation. In \textbf{ICvectorfields}, shifts to the right and up are designated as positive, whereas shifts to the left and down are considered negative. 

A typical implementation of DIC will define a region of interest within the input images or their corresponding matrices wherein displacement vectors are sought using a bounding box as in the \emph{DispFieldbb} function in \textbf{ICvectorfields} or using a sub-grid of equal sized regions of interest as in the \emph{DispField} function in \textbf{ICvectorfields}. Note that all of the functions in \textbf{ICvectorfields} that use DIC or variations of it, translate displacement or velocities in terms of pixel shifts to the spatial units defined in the projection information of the original input rasters. The coordinate information required for translation of pixel shifts to the correct spatial units is obtained using functions in the \textbf{terra} R package \citep{terra}. 

\subsection*{Extending DIC to quantify persistent movement}

\begin{figure}[htbp]
  \centering
  \includegraphics[height = 8cm, width = 14cm]{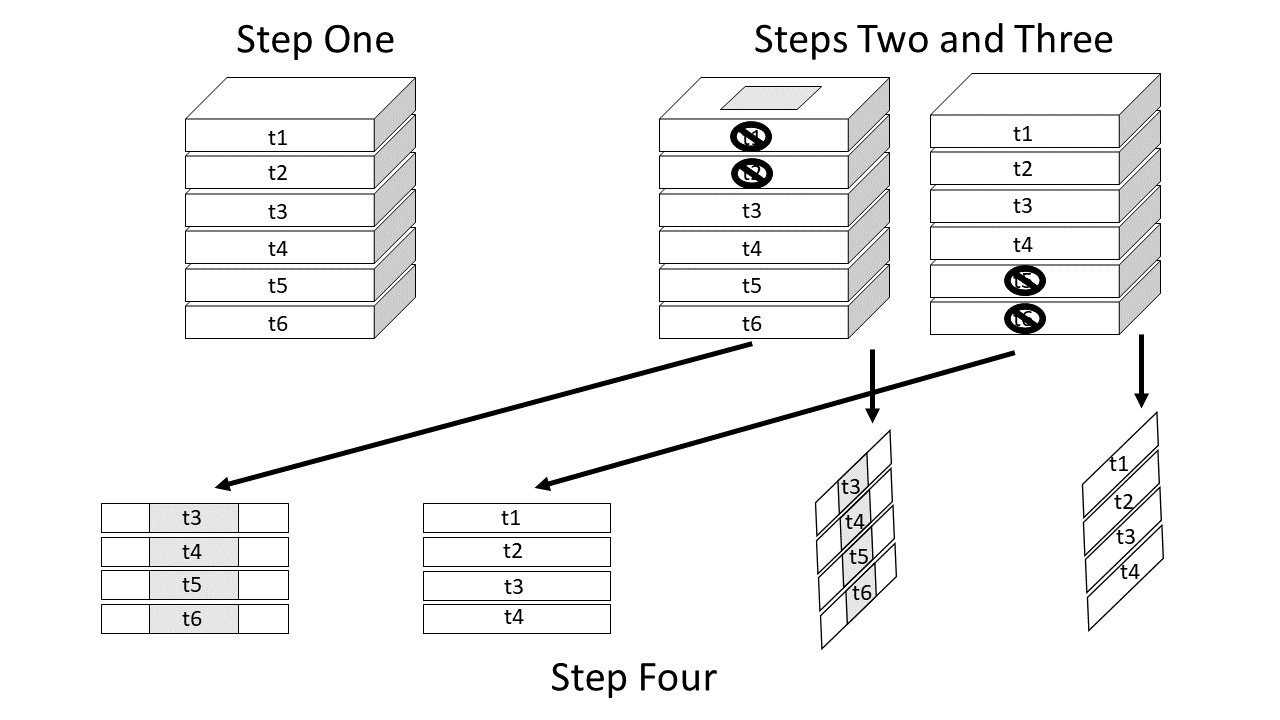}
  \caption{The STIC algorithm: The input array in step one is a raster stack of images in which each image layer represents the phenomenon of interest in planar space at a different time instance. In step two, the input array is duplicated and based on a user specified time lag, layers are removed from the top of one array and the bottom of the other. In addition, a region of interest is defined in one of the duplicate arrays represented by the grey shaded region at the top of the prism on the left. In step four the rows are dimension reduced by averaging along one of the axes (either rows or columns). This produces a pair or row-averaged matrices and a pair of column-averaged matrices that are analyzed using cross-correlation to estimate orthogonal velocity vectors.}
  \label{fig:1}
\end{figure}

In applications of DIC in earth systems with persistent directional flows that influence movement, it is valuable to determine directional movement of populations of interest that persist for more than one time step. In such situations, a spatiotemporal array of images with two space dimensions and one time dimension is required. Often these can be formulated as stacks of raster images, with each layer in the stack representing spatially referenced observations for one time step (step one in Figure~\ref{fig:1}). A variation of DIC which I call Spatiotemporal Image Correlation (STIC) permits estimation of persistent directional movement in terms of orthogonal velocity vectors.

In STIC, the three dimensional array is first lagged by duplicating it and then removing an integer number of layers from the top of one duplicate and the bottom of the other (steps two and three in Figure~\ref{fig:1}). The integer lag is user defined and serves to minimize estimates of zero movement which always would occur in the absence of a lag. To differentiate the duplicate lagged arrays, I will refer to the first as the reference array, and the second as the lagged array. Regions of interest in the reference array are selected and locations outside the region of interest in the reference array are assigned values of zero (grey shaded region in steps two and three in Figure~\ref{fig:1} represent regions of interest). The reference array and the lagged array are then dimension reduced by averaging along rows to obtain one pair of two-dimensional matrices and by averaging along columns to obtain a second pair of two dimensional matrices (step 4 in Figure~\ref{fig:1}). The first pair of matrices comprises row averaged reference and lagged matrices. The second pair of matrices comprises column averaged reference and lagged matrices. Each matrix in the two pairs has one space dimension and one time dimension.

Cross-correlation between the pairs of reference and lagged matrices is then computed as described for DIC. Recall that one dimension of each of the row or column-averaged matrices is spatial while the other is temporal, which enables calculation of two orthogonal velocity vectors based on space shifts and time shifts obtained by application of DIC:

\begin{eqnarray}
v_x &= s_x/s_{tx}, s_{tx} \neq 0\\
v_y &= s_y/s_{ty}, s_{ty} \neq 0
\end{eqnarray}

\noindent in which $v_x$ and $v_y$ are velocity in the horizontal and vertical directions, $s_x$ and $s_y$ are shifts in the horizontal and vertical direction, $s_{tx}$ is the time shift that corresponds to spatial shifts in the horizontal direction, and $s_{ty}$ is the time shift that corresponds to spatial shifts in the vertical direction. Note that due to the time shift, the user-defined time lag does not necessarily pre-determine the denominator of the orthogonal velocity vectors. 

\subsection*{Spatially variable velocities}

When the magnitudes of movement velocities are highly spatially variable, a single time lag is not optimal for quantifying orthogonal velocity vectors. For these scenarios a variation on the STIC algorithm called STIC+ allows the user to specify a maximum time lag. The algorithm then repeats the steps described for STIC for each integer time lag from one to the maximum time lag. For each repetition and each location of interest, the total velocity magnitude (speed) is calculated as 

\begin{equation}
|v| = \sqrt{v_x^2 + v_y^2}. \label{eq:8}
\end{equation}

\noindent For each region of interest, the horizontal and vertical velocity vectors are determined by the time lag STIC calculation that maximizes equation \eqref{eq:8}. 

A summary table describing which functions in \textbf{ICvectorfields} use each of the algorithms described above is provided (Table \ref{tab:1}). Two functions in \textbf{ICvectorfields} use a standard implementation of DIC similar to that described by \citet{Anuta1970}, two functions use the STIC extension and two functions use the STIC+ extension (Table \ref{tab:1}).

\begin{table}
  \centering
  \begin{tabular}{ l l l l l}
    \toprule
    function & algorithm & Images & ROI & Velocities\\
    \midrule
    \emph{DispField} & DIC & 2 &  grid &  variable or not\\
    \emph{DispFieldbb} & DIC  & 2 & bounding box & variable or not\\ 
    \emph{DispFieldST} & STIC & 3+ & grid & less variable\\ 
    \emph{DispFieldSTbb} & STIC & 3+ & bounding box & less variable\\ 
    \emph{DispFieldSTall} & STIC+ & 3+ & grid & more variable\\ 
    \emph{DispFieldSTbball} & STIC+ & 3+ & bounding box & more variable\\ 
    \bottomrule
  \end{tabular}
  \caption{\textbf{ICvectorfields} functions, algorithms, and use contexts to facilitate decisions on which function is most applicable. ROI stands for region of interest, which is defined either using a grid or a bounding box, Velocities refers to whether the magnitudes of velocities in the vector field are presumed to be spatially variable or not.}
  \label{tab:1}
\end{table}

\section*{Application}

In this section I demonstrate the use of DIC and extensions implemented in \textbf{ICvectorfields} \citep{ICvf} using an example in which data were simulated based on a partial differential equation and using the classic larch budmoth defoliation data that are embedded in the \textbf{ncf} package \citep{ncf}. The data from the simulated example are embedded in \textbf{ICvectorfields}. For visualization of results, the demonstrations call functions in several R packages: These include \textbf{ggplot2} \citep{ggplot2}, \textbf{ggnewscale} \citep{ggnewscale}, \textbf{metR} \citep{metR}, and \textbf{terra} \citep{terra}.

\subsection*{Demonstration using simulated data}

The model used to simulate data to test and demonstrate the functionality of \textbf{ICvectorfields} is a convection-diffusion equation, which is a partial differential equation with terms for diffusion, advection, and reaction:

\begin{equation}
\frac{\partial u}{\partial t} = \nabla \cdot \left(D \nabla u \right) - \nabla \cdot \left( \mathbf{v} u \right) + r u, \label{eq:9}
\end{equation}

\noindent in which $r$ is the growth rate with units of per unit time, $D$ is the diffusion coefficient with squared spatial units per unit time, $\mathbf{v}$ is the bivariate orthogonal velocity vector in units of space per unit time, $\nabla$ represents the gradient, and $\nabla \cdot$ represents divergence. The orthogonal velocity vector is spatially dependent in the simulations that follow:

\begin{equation}
\mathbf{v} = \begin{cases} \label{eq:10}
			(0, 0.2), & x < 0, y \geq 0\\
      (0.2, 0), & x \geq 0, y \geq 0\\
      (0, -0.2), & x \geq 0, y < 0\\
      (-0.2, 0), & x < 0, y < 0
		 \end{cases}
\end{equation}

\noindent Note that by convention in equations \eqref{eq:9} and \eqref{eq:10}, movement to the right and up has a negative sign, and movement to the left and down has a positive sign. This is the opposite convention used in \textbf{ICvectorfields}. Note that the discontinuities at $x = 0$ and $y = 0$ in the advection term in equation \eqref{eq:10} create strange model behaviours once concentration reaches $x = 0$ or $y = 0$, and so the model was simulated for only 6 time steps to minimize encounters with these axes. Parameter values for the diffusion coefficient and the growth rate were $D = (0.01, 0.01)$ squared spatial units per unit time and $r = 0.5$ per unit time.

The model in equations \eqref{eq:9} and \eqref{eq:10} was simulated using the \textbf{ReacTran} R package \citep{ReacTran}, using a finite differencing scheme with backward differencing on a square domain of 202 cells in each direction, each with a width of 0.049 spatial units. The initial condition was a concentration of one units per arbitrary unit of volume in the centre of each quadrant of the spatial domain. Boundary conditions were zero flux (reflecting) on all four sides of the spatial domain. The simulation data are saved in table format within \textbf{ICvectorfields}.

The data are imported and then converted from table format to a raster stack using the \emph{RastStackData} helper function. They can then be visualized as rasters as shown below.

\begin{verbatim}
# import simulated data
data(SimData, package = "ICvectorfields")

# convert to raster stack
SimStack <- ICvectorfields::RastStackData(SimData)

# confirming dimension
dim(SimStack)

# plotting
layout(matrix(1:6, 2, 3, byrow = TRUE))
layout.show(6)
terra::plot(SimStack[[1]], legend = FALSE, main = "t1")
terra::plot(SimStack[[2]], legend = FALSE, main = "t2")
terra::plot(SimStack[[3]], legend = FALSE, main = "t3")
terra::plot(SimStack[[4]], legend = FALSE, main = "t4")
terra::plot(SimStack[[5]], legend = FALSE, main = "t5")
terra::plot(SimStack[[6]], legend = FALSE, main = "t6")
\end{verbatim}

\begin{figure}[htbp]
  \centering
  \includegraphics[height = 8cm, width = 14cm]{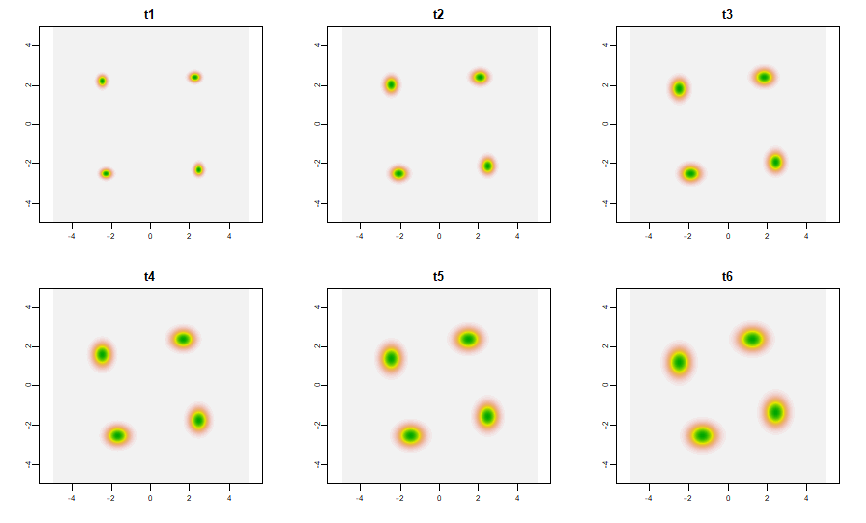}
  \caption{Visualization of simulation data for six time steps. The initial condition at t0 is not shown. Green colours represent the highest concentrations.}
  \label{fig:2}
\end{figure}

The figure showing the simulated spatiotemporal dynamics is identical to Figure~\ref{fig:2}.

To analyze displacement based on a pair of images, I chose to use the standard implementation of DIC in the 
\emph{DispField} function of \textbf{ICvectorfields}. The first two arguments of the \emph{DispField} function are the input rasters. The first input raster is treated as the reference image and the second is treated as the shifted image. In this case, the first image is the raster layer corresponding to the first time step (t1) and the second image is the raster layer corresponding to the sixth time step (t6). The function selects regions of interests based on a grid of dimensions given in the \emph{factv1} and \emph{facth1} arguments, which represent to number of rows and columns in each sub-grid. Sub-grids start in the upper left corner and as many sub-grids as fit within the original domain are constructed. In the code below, sub-grids are $101 \times 101$, which is approximately the size of one quarter of the original spatial domain of the simulation. The restricted argument is by default set to \emph{FALSE}. In that case, the DIC algorithm cross-correlates each region of interest in the first image with the entirety of the second image. When \emph{restricted = TRUE} The algorithm cross correlates both images only within the region of interest. If the user has reason to believe that movement is predominantly occurring within sub-grids the \emph{restricted = TRUE} option has the added benefit of speeding up computation.

\begin{verbatim}
VFdf1 <- DispField(SimStack[[1]], SimStack[[6]], factv1 = 101, facth1 = 101, 
                    restricted = TRUE)
VFdf1
\end{verbatim}

The output of \emph{DispField} is in data table format. Because the data table is small, I've printed it to the console in the code example above. For ease of viewing in this paper, the output is duplicated in Table \ref{tab:2}.

\begin{table}
  \centering
  \begin{tabular}{ c c c c c c c c c c }
    \toprule
    row & col & frowmin & frowmax & fcolmin & fcolmax & centx & centy & dispx & dispy\\
    \midrule
    51  & 51  & 1   & 101 & 1   & 101 & -2.5 &  2.5 &  0.00 & -0.98\\
    152 & 51  & 102 & 202 & 1   & 101 & -2.5 & -2.5 &  0.98 &  0.00\\ 
    51  & 152 & 1   & 101 & 102 & 202 &  2.5 &  2.5 & -0.98 &  0.00\\
    152 & 152 & 102 & 202 & 102 & 202 &  2.5 & -2.5 &  0.00 &  0.98\\
    \bottomrule
  \end{tabular}
  \caption{\textbf{ICvectorfields} output from a call of the \emph{DispField} function using simulated data. The table is a duplicate of the data table returned after calling the function except that the headings for the first two columns have been shortened from rowcent to row and from colcent to col so that the table fits the page. In addition centx and centy have been rounded to a single decimal place and dispx and dispy have been rounded to two decimal places.}
  \label{tab:2}
\end{table}

The directions of movement coincide with the directions of advection in the simulation with movement downwards in the upper left quadrant (first row of Table \ref{tab:2}), movement to the right in the lower left quadrant (second row of Table \ref{tab:2}), movement to the left in the upper right quadrant (third row of Table \ref{tab:2}), and upwards movement in the lower right quadrant (fourth row of Table \ref{tab:2}). Speed of movement can be computed by dividing displacement by the number of time steps that passed $0.98/5 = 0.196$, which is slightly slower than the simulated advection speed of 0.2 spatial units per time step. The discrepancy is likely due to the blurring effect of diffusion in the partial differential equation (equation \eqref{eq:9}).

In situations where the speed is constant, velocity can be estimated from pairs of images as I have demonstrated above. However, the \emph{DispFieldST} algorithm is designed to return orthogonal velocity vectors and so for confirmation purposes, I demonstrate it below:

\begin{verbatim}
VFdf2 <- DispFieldST(SimStack, lag1 = 1, factv1 = 101, facth1 = 101, restricted = TRUE)
VFdf2
\end{verbatim}

The data table that is printed to the console after running the code above looks similar to the data table duplicated in Table \ref{tab:2} except that under the heading dispx and dispy the algorithm returns horizontal and vertical velocities rather than displacement vectors. The directions of movement are the same as those shown in Table \ref{tab:2}, but the speed is 0.196 spatial units per unit time as previously estimated.

The vector field and the raw data can be visualized simultaneously using plotting functionality in \textbf{ggplot2} with extensions in the \textbf{ggnewscale} and \textbf{metR} packages.

\begin{verbatim}
SimVF = ggplot() +
  xlim(c(-5, 5)) +
  ylim(c(-5, 5)) +
  geom_raster(data = SimData,
              aes(x = xcoord, y = ycoord, fill = t1)) +
  scale_fill_gradient(low = "white", high = "blue", na.value = NA) +
  new_scale("fill") +
  geom_raster(data = SimData,
              aes(x = xcoord, y = ycoord, fill = t6), alpha = 0.5) +
  scale_fill_gradient(low = "white", high = "red", na.value = NA) +
  geom_vector(data = VFdf2, 
              aes(x = centx, y = centy, 
                  mag = Mag(dispx, dispy), 
                  angle = Angle(dispx, dispy))) + 
  theme_bw()
SimVF
\end{verbatim}

\begin{figure}[htbp]
  \centering
  \includegraphics[height = 8cm, width = 10cm]{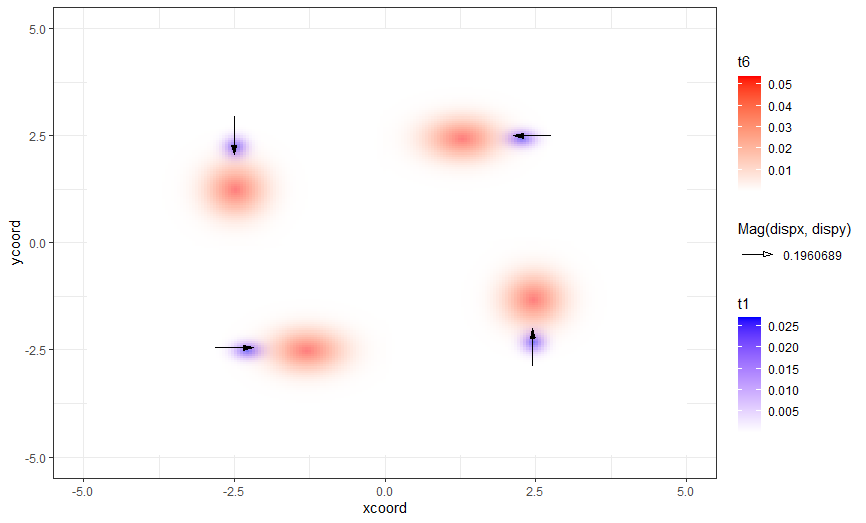}
  \caption{Vector field for radial movement simulated using a convection-diffusion equation. the orthogonal velocity vectors are estimated using the \emph{DispFieldST} function in \textbf{ICvectorfields}.}
  \label{fig:3}
\end{figure}

The resulting figure is duplicated in Figure~\ref{fig:3}. The velocity vectors in the vector field (Figure~\ref{fig:3}) are consistent with the simulated advection vectors, although they slightly underestimate movement speed.

Before proceeding to the next demonstration I will illustrate one of the potential pitfalls of estimating movement based on cross-correlation. If the argument of the \emph{DispFieldST} function is left its default \emph{restricted = FALSE} configuration, the algorithm will search the entire domain for shifts that maximize cross-correlation. Because the simulations in each quadrant of the spatial domain are quite similar, cross-correlation is in fact maximized by shifts that cross quadrants, even though simulated movement was not that large. Therefore, calling \emph{DispFieldST} with \emph{restricted = FALSE} produces incorrect output (Table \ref{tab:3}): The simulated advection speed is not at all close to the estimated maximum orthogonal advection speed of 3.87 spatial units per unit time.

\begin{table}
  \centering
  \begin{tabular}{ c c c c c c c c c c }
    \toprule
    row & col & frowmin & frowmax & fcolmin & fcolmax & centx & centy & dispx & dispy\\
    \midrule
    51  & 51  & 1   & 101 & 1   & 101 & -2.5 &  2.5 &  0.10 & -3.87\\
    152 & 51  & 102 & 202 & 1   & 101 & -2.5 & -2.5 &  3.87 &  0.10\\ 
    51  & 152 & 1   & 101 & 102 & 202 &  2.5 &  2.5 & -3.87 & -0.10\\
    152 & 152 & 102 & 202 & 102 & 202 &  2.5 & -2.5 & -0.10 &  3.87\\
    \bottomrule
  \end{tabular}
  \caption{\textbf{ICvectorfields} output from a call of the \emph{DispFieldST} function using simulated data. This call is meant to demonstrate a potential pitfall in using the cross-correlation approach because when \emph{restricted = FALSE}, the algorithm finds positive cross-correlations that are not caused by movement. The table is a duplicate of the data table returned after calling the function except that the headings for the first two columns have been shortened from rowcent to row and from colcent to col so that the table fits the page. In addition centx and centy have been rounded to a single decimal place and dispx and dispy have been rounded to two decimal places.}
  \label{tab:3}
\end{table}

\subsection*{Demonstration using larch budmoth data}

Larch budmoths are lepidopteran defoliators that exhibit periodic outbreaks every 8 to 9 years in the European Alps \citep{Bjornstad2002}. The larch budmoth data originated from survey information collected by the forest administrative agencies of France, Italy, Switzerland, and Austria from 1961 to 1998. The data record the presence of defoliation by larch budmoth caterpillars within 1 $\times$ 1 km grid cells (a binary variable). These data were aggregated up to 20 $\times$ 20 km grid cells so that records at this spatial scale were population proxies for larch budmoth caterpillar abundance \citep{Bjornstad2002} based on the assumption that defoliation damage is proportional to the abundance of the causal agents. Grid cells were excluded from the data set if they exhibited less than one percent defoliation or if more than ninety percent of years in which data were collected at that location exhibited no defoliation by larch budmoth \citep{Bjornstad2002}. The larch budmoth defoliation data exhibit directional traveling wave-trains that travel from the southwest to the northeast along the European Alps \citep{Bjornstad2002}. These data are embedded in the \textbf{ncf} R package.

After loading the \textbf{ncf} library as well as \textbf{ICvectorfields}, the data can be loaded, converted to a raster stack and visualized as follows:

\begin{verbatim}
# import larch budmoth data
data(lbm, package = "ncf")

# convert to raster stack
LBMStack <- ICvectorfields::RastStackData(lbm)

# confirming dimension
dim(LBMStack)

# visualizing
layout(matrix(1:6, 2, 3, byrow = TRUE))
layout.show(6)
terra::plot(LBMStack[[1]], legend = FALSE, main = "1961")
terra::plot(LBMStack[[2]], legend = FALSE, main = "1962")
terra::plot(LBMStack[[3]], legend = FALSE, main = "1963")
terra::plot(LBMStack[[4]], legend = FALSE, main = "1964")
terra::plot(LBMStack[[5]], legend = FALSE, main = "1965")
\end{verbatim}

\begin{figure}[htbp]
  \centering
  \includegraphics[height = 8cm, width = 14cm]{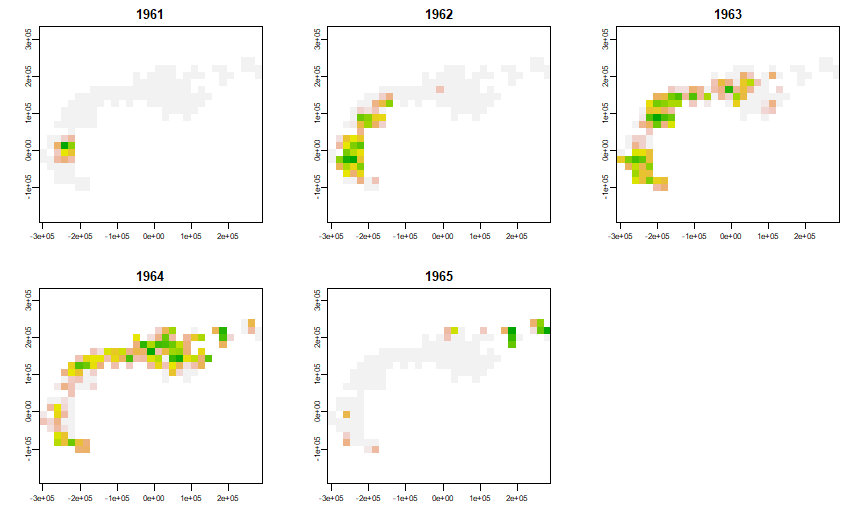}
  \caption{The first five years of the larch budmoth defoliation data included in \textbf{ncf}. Green colours represent the highest level of defoliation.}
  \label{fig:4}
\end{figure}

This code plots the first five years of the data set (Figure~\ref{fig:4}), which show a standard progression of outbreaks from the southwest corner of the Alps to the northeast. This pattern repeats relatively regularly every 8 to 10 years in the data set.

The study region covers a large geographic area and so it is likely the population movement speeds vary geographically. For this reason, I elected to use the STIC+ algorithm to analyze the data using \emph{DispFieldSTall}. In the code below I analyze the first 23 years of the time series (1961 to 1983) as defoliation patterns from 1984 to 1998 are less regular.

\begin{verbatim}
VFdf3 <- DispFieldSTall(LBMStack[[1:23]], lagmax = 3, factv1 = 3, facth1 = 3, restricted = FALSE)

LBMVF1 = ggplot() +
  geom_tile(data = lbm,
              aes(x = x, y = y, 
                  fill = X1962)) +
  scale_fill_gradient(low = "white", high = "blue", na.value = NA) +
  new_scale("fill") +
  geom_tile(data = lbm,
              aes(x = x, y = y, fill = X1964), alpha = 0.5) +
  scale_fill_gradient(low = "white", high = "red", na.value = NA) +
  geom_vector(data = VFdf3, 
              aes(x = centx, y = centy, 
                  mag = Mag(dispx, dispy), 
                  angle = Angle(dispx, dispy))) + 
  theme_bw()
LBMVF1
\end{verbatim}

\begin{figure}[htbp]
  \centering
  \includegraphics[height = 8cm, width = 10cm]{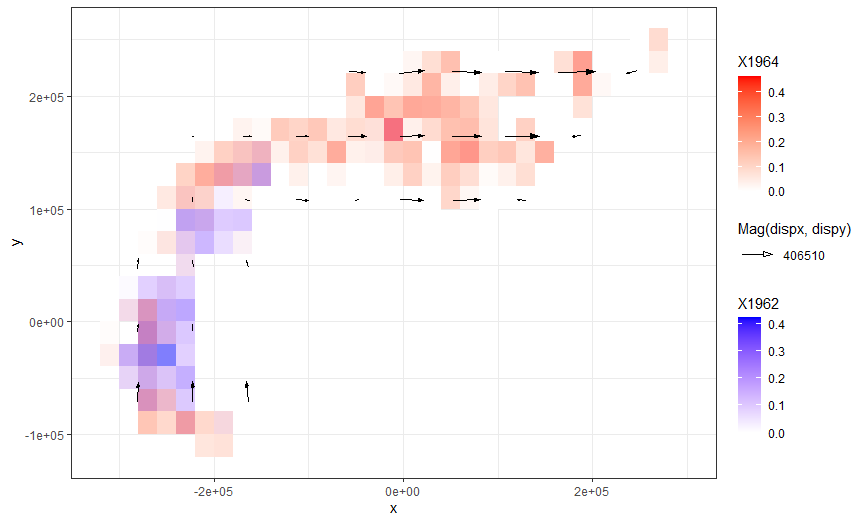}
  \caption{Vector field for Larch Budmoth persistent movement. The orthogonal velocity vectors are estimated using the \emph{DispFieldSTall} function in \textbf{ICvectorfields}. Blue colours show the locations and intensities of defoliation in 1962 and red colours show the locations and intensities of defoliation in 1964. Vectors have their own scale that is distinct from the scale of the map.}
  \label{fig:5}
\end{figure}

Calling \emph{DispFieldSTall} returns a data frame object that is convenient for plotting the vector field. The vector field reveals that moth movement is to the north on the southwestern side of the Alps and then to the east on the northern side of the Alps (Figure~\ref{fig:5}). It also reveals deceleration as outbreaks turn the corner and then acceleration as outbreaks move eastward (Figure~\ref{fig:5}).

The average speed of larch budmoth movement can be computed from the data frame output of \emph{DispFieldSTall} as follows:

\begin{verbatim}
VFdf3$speed <- sqrt((VFdf3$dispx^2) + VFdf3$dispy^2)

# subsetting to remove locations where speed is zero
VFdf4 <- subset(VFdf3, speed > 0)

# computing mean, standard deviation and dimension of data frame
# to obtain sample size
mean(VFdf4$speed)
sd(VFdf4$speed)
dim(VFdf4)

# upper and lower Wald-type 95 percent confidence interval on average speed
mean(VFdf4$speed)/1000 + qt(0.975, dim(VFdf4)[1] - 1)*sd(VFdf4$speed)/1000/sqrt(dim(VFdf4)[1] - 1)
mean(VFdf4$speed)/1000 + qt(0.025, dim(VFdf4)[1] - 1)*sd(VFdf4$speed)/1000/sqrt(dim(VFdf4)[1] - 1)
\end{verbatim}

Using the approach above, the average movement speed is estimated as $176 \pm 43$ $\text{km}(\text{Yr})^{-1}$, an estimate that is less than the previous speed estimates for northeastern spread of 220 $\text{km}(\text{Yr})^{-1}$ \citep{Bjornstad2002} and 254 $\text{km}(\text{Yr})^{-1}$ \citep{Johnson2004}. The difference between estimates in the literature and estimates produced here are likely due to the direction of movement. The vectors in the larch budmoth vector field point predominantly north and east (Figure~\ref{fig:5}). In other words they are orthogonal. In contrast the movement speeds estimated by \citet{Bjornstad2002} and \citet{Johnson2004} are projected along lines that run to the northeast. A simple application of geometry reveals that an average speed of 176 $\text{km}(\text{Yr})^{-1}$ in the north and east directions corresponds to an estimated speed of 249 $\text{km}(\text{Yr})^{-1}$ in the northeastern direction (Pythagorean theorem $\sqrt{176^2 + 176^2}$). This estimate is consistent with prior speed estimates for larch budmoth population movement \citep{Bjornstad2002, Johnson2004}.

\section*{Summary}

The \textbf{ICvectorfields} R package implements standard Digital Image Correlation algorithms in addition to a novel extension that permits estimation of orthogonal velocities of persistent movement in series of three or more images. Here I demonstrate the usefulness of DIC and the extension implemented in \textbf{ICvectorfields} in a new arena: Whereas DIC is often applied in engineering and materials science to quantify the effects of force application on materials \citep{Sutton2009}, it has not been used in landscape ecology. In this field, the approach has potential to provide new insights into how populations move across landscapes and to demonstrate the untenable nature of assumptions of homogeneity inherent in most analyses based on the traveling wave paradigm. Even when models of sufficient complexity to capture environmental heterogeneity can be used, I expect that the methods in \textbf{ICvectorfields} will be useful because they facilitate comparison between modeled and empirical population movement data as demonstrated in the partial differential equation example in this study. Approaches such as this one that estimate movement based on cross-correlation, however, have a weakness: Under certain circumstances, they are prone to finding cross-correlations that are unrelated to movement as was demonstrated in this paper. For this reason, users must exercise vigilance in interpreting the results of vector field analyses like those demonstrated herein. If possible, results should be checked against a standard or against prior published results regarding movement propensity. Nevertheless, the methods described here hold promise for exploratory analyses, hypothesis generation, and synoptic pattern analyses of population movements. 

\bibliography{ImageCorrelationVFarxiv}

\end{document}